\documentclass[preprint,12pt]{elsarticle}

\usepackage{amssymb}
\usepackage{amsmath}
\usepackage{bm}
\usepackage{float}
\usepackage{subfigure}
\usepackage{booktabs}
\usepackage{tabularx}
\makeatletter
\def\ps@pprintTitle{}
\makeatother

\begin{document}

\begin{frontmatter}



\title{Optimal Operation of Distribution Networks under Asymmetric Renewable Energy and Load Demand Uncertainties}


\author[1]{Zhisheng Xiong}
\author[2]{Bo Zeng}
\author[1]{Peter Palensky}
\author[1]{Pedro P. Vergara}

\affiliation[1]{organization={Intelligent Electrical Power Grids (IEPG) group, Delft University of Technology},
            addressline={2628 CD},
            city={Delft},
            country={Netherlands}}

\affiliation[2]{organization={Department of Industrial Engineering and Department of Electrical and Computer Engineering, University of Pittsburgh},
            addressline={PA 15106},
            city={Pittsburgh},
            country={USA}}

\begin{abstract}
To develop an optimal operational scheme for distribution networks capable of addressing asymmetric uncertainties associated with renewable energy and load demands, this paper presents a confidence level-based information gap decision theory (CL-IGDT) framework. Building on IGDT, the proposed framework utilizes the confidence level to capture the asymmetric characteristics of uncertainties and maximize the risk-averse capability of the solution in a probabilistic manner. To facilitate such probabilistic consideration, the imprecise Dirichlet model is employed to construct the ambiguity sets of uncertainties. Consequently, a two-stage robust optimal operation model for distribution networks using CL-IGDT is developed. An iterative method is proposed to solve the model and determine the upper and lower bounds of the objective function. Case study demonstrates that the proposed approach yields a more robust and statistically optimized solution with required accuracy compared to existing method, contributing to a reduction in first-stage cost by 0.84\%, second-stage average cost by 6.7\%, and significantly increasing the reliability of the solution by 8\%.
\end{abstract}




\end{frontmatter}

\section*{Nomenclature}

\begin{tabularx}{\linewidth}{@{}lX@{}}
\multicolumn{2}{@{}l}{\raggedright \underline{\textbf{\emph{Sets}}}} \\[2pt]
    $\mathcal{N}$ & Set of nodes of the distribution network \\
    $\mathcal{L}$ & Set of lines \\
    $\mathcal{G}$ & Set of nodes where DG units are connected, $\mathcal{G} \subset \mathcal{N}$ \\
    $\mathcal{B}$ & Set of nodes where ESS units are connected, $\mathcal{B} \subset \mathcal{N}$ \\
    $\mathcal{T}$ & Set of time periods \\[6pt]
    
\multicolumn{2}{@{}l}{\raggedright \underline{\textbf{\emph{Indices}}}} \\[2pt]
    $ij$ & Line $ij \in \mathcal{L}$ \\
    $i, j$ & Nodes $i, j \in \mathcal{N}$ \\
    $t$ & Time period $t \in \mathcal{T}$ \\[6pt]

\multicolumn{2}{@{}l}{\raggedright \underline{\textbf{\emph{Parameters}}}} \\[2pt]
    $c_i$, $\hat{c}_i$ & First-/second-stage DG operational cost coefficients \\
    $d_t$, $\hat{d}_t$ & First-/second-stage electricity purchase cost coefficients \\
    $\overline{e}_i$, $\underline{e}_i$ & Maximum/minimum energy capacity of ESSs \\
    $\overline{p}^{B+}$, $\overline{p}^{B-}$ & Maximum charging/discharging power of ESSs \\
    $\eta_B$ & Charging/discharging efficiency of ESSs \\
    $\overline{p}_i^{G}$, $\underline{p}_i^{G}$ & Maximum/minimum active power of DGs \\
    $\overline{\hat{p}}_i^{G}$, $\underline{\hat{p}}_i^{G}$ & Maximum/minimum recourse output of DGs \\
    $\overline{\hat{q}}_i^{G}$, $\underline{\hat{q}}_i^{G}$ & Maximum/minimum reactive power output of DGs \\
    $\overline{p}^{l}$ & Maximum power flow limit \\
    $\overline{v}$, $\underline{v}$ & Maximum/minimum limit of $\hat{v}_{i,t}$ \\
    $r_{ij}$, $x_{ij}$ & Resistance and reactance of line $ij$ \\
    $RU_i$, $RD_i$ & Ramp up/down limits of DGs \\
    $p^{PV,cap}$ & Installed capacity of solar panels \\
    $\Delta t$ & Duration of one time step \\[6pt]

\multicolumn{2}{@{}l}{\raggedright \underline{\textbf{\emph{Continuous Variables}}}} \\[2pt]
    $e_{i,t}$ & Stored energy of ESS \\
    $\widehat{p}_{i,t}^{G}$ & Active power output of DGs after recourse action \\
    $p_{t}^{S}$, $\hat{p}_{t}^{S}$ & First-/second-stage purchased power from main grid \\
    $p_{ij,t}$, $\hat{p}_{ij,t}$ & First-/second-stage active power flow \\
    $q_{ij,t}$, $\hat{q}_{ij,t}$ & First-/second-stage reactive power flow \\
    $v_{i,t}$, $\hat{v}_{i,t}$ & First-/second-stage squared bus voltage magnitude \\
    $\delta_{i,t}$ & Power factor angle \\
\end{tabularx}

\section{Introduction}
\label{Introduction}
Distribution networks (DNs) are increasingly integrating high shares of renewable energy sources (RESs). The intermittent and volatile nature of RESs, along with the uncertainty in load demands, can affect both the economics and security of the networks \cite{impact_RESs}. Failure to appropriately address uncertainty may increase operational costs due to actions such as load shedding, RESs curtailment, and the need for excessive reserves. Moreover, potential risks such as voltage violations and line overloading may occur. These significant challenges to the optimal operation of DNs may undermine the transition to more sustainable energy systems \cite{impact_RESs2}. Accordingly, numerous studies have been conducted to manage the uncertainties associated with RESs and load demands, aiming to develop robust operational strategies that enhance the cost-effectiveness and reliability of DNs.

Most research on the optimal operation of DNs under uncertainty relies on varying levels of information about the random variables: complete, partial, or none at all. Ref.~\cite{TS_CC} models the wind power loss and line overloading as chance constraints using a truncated normal distribution, establishing a two-stage stochastic programming (SP) model for unit commitment. Ref.~\cite{SP_ED} presents a two-stage stochastic dynamic economic dispatch model that effectively addresses wind power uncertainty. This model improves cost efficiency and reliability by pre-dispatching generator output to avoid network congestion in the first stage and re-dispatching resources after the realization of the wind scenarios. Ref.~\cite{P_Q_stochastic} incorporates multiple correlations of RESs through probability distribution functions (PDFs) and scenario analysis into a multi-time scale SP framework. By managing slow- and fast-response resources to coordinate active and reactive power, the economical performance and the secure operation of the system are ensured under RESs uncertainty. SP features the need for PDFs or uncertain scenarios, making its performance closely tied to the estimation of distributional information or the number of scenarios, which can be well guaranteed with an incredibly large amount of historical data.

Robust optimization (RO) uses the boundary information of uncertainties, making it a practical choice for the optimal operation of DNs \cite{RO_Introduction}. However, solutions derived from RO tend to be conservative because they focus solely on the worst-case scenario within a predefined uncertainty set, ignoring any information on the uncertainties. To mitigate the conservatism, some researchers employed partial information to refine the uncertainty set. Ref.~\cite{Data-Adaptive} searches the extreme scenarios from the historical data of RESs to redesign the uncertainty sets, yielding a less conservative yet robust solution for determining the tap ratio of transformers and the capacity of switching capacitors. Ref.~\cite{ReconRO} reconstructs the uncertainty set of wind power based on the historical data of forecasted errors, aiming to reduce conservatism in economic dispatch. Beyond RO, distributionally robust optimization (DRO) incorporates available probability information on uncertainties to minimize operational cost over the worst-case distribution within the ambiguity sets constructed from historical data. Ref.~\cite{two-stage_modeling_DRO} introduces a moment-based DRO to characterize RESs uncertainty for real-time power dispatch in DNs. Ref.~\cite{Wasserstein} constructs an ambiguity set of wind power using Wasserstein-based DRO, followed by a chance constraint-based DRO model for reactive power dispatch. Ref.~\cite{idm_Li_Peng} introduces the imprecise Dirichlet model (IDM) to construct the ambiguity set of wind power distribution, with an expected risk indicator to identify the worst-case distribution, aiming to balance operational costs and the risk costs in real-time energy dispatch.

It is well-established that increasing the uncertainty set (ambiguity set) enhances robustness at the cost of reduced economic performance. Several critical challenges thus arise in the optimization process, such as the size of the uncertainty set, the amount of operational cost budget allocated, as well as the trade-off between robustness and economic performance. Many of the aforementioned approaches fail to simultaneously address these challenges effectively. Information gap decision theory (IGDT) provides a robust decision-making framework for handling severe uncertainties without requiring any information or a predefined uncertainty set \cite{igdt_Ben_Haim}. Risk-averse based-IGDT seeks to optimize the uncertainty set (representing risk-averse capability) within a preset financial budget, which is particularly useful when there is a clear target. For instance, given an operational budget, an IGDT-based energy management system is proposed for islanded microgrids in \cite{IGDT_frequency_excursions} to maximize the allowable range of load and RESs fluctuations while managing the system frequency excursion. Ref.~\cite{resiliency-oriented_IGDT} applies IGDT to identify robust operational scheme for resiliency-oriented DNs, addressing uncertainties in RESs, load demand, etc., while adhering to a certain budget. Ref.~\cite{w_igdt} imcorporates the penetration level of RESs at different time periods into the uncertainty modeling framework, and proposes a weighted-IGDT to handle wind power uncertainty in microgrid. To account for the electricity price uncertainty and their correlations, the ellipsoid-bound IGDT has been utilized to model the electric vehicle aggregator scheduling in \cite{ellipsoid-IGDT}.

 The envelope-bound uncertainty modeling dominates the current IGDT studies in network operation \cite{IGDT_frequency_excursions, resiliency-oriented_IGDT, w_igdt}, as well as in network planning \cite{igdt_network_planning}, electricity market offering strategy \cite{igdt_electricity_market}, etc., due to its simplicity and capability to handle forecasted types of uncertainty sources \cite{igdt_Ben_Haim}. However, the uncertainty modelling of IGDT does not incorporate any information on uncertain parameters, and the real-valued symmetric uncertainty sets limit the approach to capture the asymmetric characteristics of uncertainties. The asymmetry of uncertainties primarily refers to scenarios that deviate significantly from the expected values yet have a higher probability of occurrence than closer scenarios, as commonly observed in the heavy tails of PDFs. For instance, scenarios with low RESs generation and high load demand, significantly deviating from expected values, are uncommon yet still possible and may be excluded from the uncertainty sets. These scenarios can have substantial impacts when they occur. However, many existing uncertainty modelling approaches fail to account for these significant scenarios probabilistically, or neglect to incorporate the desired operational budget when defining uncertainty sets. While SP considers the probability of scenarios, as previously discussed, its accuracy is highly dependent on the estimation of distributional information and the size of the dataset.

 Given the advantages and disadvantages of the IGDT approach and its widespread utilization of envelope-bound uncertainty modeling, it is essential to seek potential enhancements to improve its applicability in DNs operations. Therefore, we propose a novel uncertainty modelling approach that utilizes confidence level to capture the asymmetric characteristics of uncertain parameters and maximize risk-averse capability in a probabilistic manner within the IGDT framework, while adhering to a desired budget. The IDM is employed to construct the ambiguity sets, leveraging available data and mitigating the dependency on accurate probability distributions. Consequently, the confidence level-based IGDT (CL-IGDT) framework is developed for the two-stage robust optimal operation of DNs. An iterative method is proposed to solve the model and determine the upper and lower bounds of the objective function. Overall, the main contributions of this paper are as follows:

\begin{itemize}
	\item A two-stage DNs optimal operation model is constructed using CL-IGDT. In the proposed framework, the confidence interval replaces the real-valued symmetric interval of IGDT, which captures the asymmetric characteristics of the uncertainties associated with RESs and load demands. Additionally, the newly defined objective function ensures the robustness of the solution by describing better the risk-averse capability in a probabilistic way, compared with the traditional IGDT.
	\item IDM is utilized to construct the ambiguity sets for RESs and load demands, and integrated into the proposed optimization model. This method reduces reliance on precise probability distributions and leverages the available data at hand. 
    \item To solve the integrated model, an iterative method is employed to determine the upper and lower bounds 
    of the objective function. The iterative process continues until the objective value converges to a specified accuracy, at which point the solution is deemed sufficiently precise.
\end{itemize} 

\section{Mathematical Formulation}
\label{Mathematical Formulation}

The necessity for a two-stage operation model arises from the challenges of multi-time scale control in DNs \cite{two-stage_modeling_DRO}. Consequently, this section introduces an optimization model for DNs optimal operation problem, consisting of an economic dispatch plan and a recourse control stage. In the first stage, the approximate active power plan is scheduled, aiming at economic dispatch under expected uncertain scenarios. In the second stage, after the actual realization of uncertain scenarios, both active and reactive power regulation are implemented to minimize resource costs and ensure system security. The active power plan serves as a baseline decision that guides second-stage adjustments, ensuring the implementations remain within an economically and operationally feasible range. Two types of distributed energy resources are considered dispatchable: distributed generators (DGs) and energy storage systems (ESSs).

\subsection{First-stage Economic Dispatch Plan}
In the first stage, decisions regarding the active power of DGs ($p_{i,t}^G$) and the charging/discharging power of ESSs ($p_{i,t}^{B+}/p_{i,t}^{B-}$) are defined based on expected forecasts for renewable energy ($\tilde{p}_{i,t}^{PV}$) and load demands ($\tilde{p}_{i,t}^{L}$), to determine the approximate active power output plan.

\subsubsection{Objective Function}

The operational costs $(\Lambda_1)$ include the cost of active power output from the DGs and the cost of electricity purchased from the upstream main grid.
\begin{equation}\label{eq:obj1}
\min \Lambda_1 = \sum_{t \in {\cal T}} \left(\sum_{i \in {\cal G}} c_i p_{i,t}^G + d_t p_t^{S}\right)
\end{equation}

\subsubsection{DGs Constraints}
The DGs constraints are composed of the active power output limits \eqref{eq:DGs_limit}, and the ramp up/down rate limits \eqref{eq:DGs_rampup}--\eqref{eq:DGs_rampdown}.
\begin{subequations} \label{eq:DGs}
\begin{align}
    \label{eq:DGs_limit}
    &\underline{p}_{i}^{G} \leq p_{i,t}^G \leq \overline{p}_{i}^{G}, \forall i \in {\cal G}, \forall t \in {\cal T} \\
    \label{eq:DGs_rampup}
    &p_{i,t}^{G} - p_{i,t-1}^{G} \leq RU_{i}, \forall i \in {\cal G}, \forall t \in {\cal T} \\
    \label{eq:DGs_rampdown}
    &p_{i,t-1}^{G} - p_{i,t}^{G} \leq RD_{i}, \forall i \in {\cal G}, \forall t \in {\cal T}
\end{align}
\end{subequations}

\subsubsection{ESSs Constraints}

The ESSs are considered to operate in power control mode, allowing for pre-scheduling of their active power output \cite{active_power_ESS}. Their operation is governed by constraints on stored energy dynamics \eqref{eq:ESSs_energy}, storage limits \eqref{eq:ESSs_limit}, and charging/discharging limits \eqref{eq:ESSs_charge}--\eqref{eq:ESSs_discharge}.
\begin{subequations}\label{eq:ESSs}
\begin{align}
    \label{eq:ESSs_energy}
    &e_{i,t} = e_{i,t-1} + \eta_{B} p_{i,t}^{B+} \Delta t - \frac{p_{i,t}^{B-}}{\eta_{B}} \Delta t, \forall i \in {\cal B}, \forall t \in {\cal T} \\
    \label{eq:ESSs_limit}
    &\underline{e}_{i} \leq e_{i,t} \leq \overline{e}_{i}, \forall i \in {\cal B}, \forall t \in {\cal T} \\
    \label{eq:ESSs_charge}
    &0 \leq p_{i,t}^{B+} \leq z_{i,t} \overline{p}^{B+}, \forall i \in {\cal B}, \forall t \in {\cal T} \\
    \label{eq:ESSs_discharge}
    &0 \leq p_{i,t}^{B-} \leq (1-z_{i,t}) \overline{p}^{B-}, \forall i \in {\cal B}, \forall t \in {\cal T}
\end{align}
\end{subequations}
where $z_{i,t}$ is the binary variable ensuring ESSs operate in only one state at a time -- charging, discharging, or stand-by.

\subsubsection{Power Flow Constraints}
Constraints \eqref{eq:power_balance} model the active power balance equations, and constraints \eqref{eq:netload} consider the net active loads $(p_{i,t})$.
\begin{subequations}
\begin{align}
    \label{eq:power_balance}
    &\sum_{hi \in \mathcal{L}} p_{hi,t} - \sum_{ij \in \mathcal{L}} p_{ij,t} = p_{i,t}, \forall i \in \mathcal{N}, \forall t \in \mathcal{T} \\
    \label{eq:netload}
    &p_{i,t} = p_{i,t}^{G} + p_{i,t}^{B+} - p_{i,t}^{B-} + \tilde{p}_{i,t}^{PV} - \tilde{p}_{i,t}^{L}, \forall i \in \mathcal{N}, \forall t \in \mathcal{T}
\end{align}
\end{subequations}

\subsubsection{Network Security Constraints}
The maximum power flow limits are enforced by:
\begin{equation}\label{eq:power_limit}
    -\overline{p}^{l} \leq p_{ij,t} \leq \overline{p}^{l}, \forall ij \in {\cal L}, \forall t \in {\cal T}
\end{equation}

\subsection{Second-stage Recourse Control Stage}
Decisions on the recourse actions involving the active and reactive power outputs of DGs ($\hat{p}_{i,t}^G$/$\hat{q}_{i,t}^G$) are defined based on the real realizations of renewable energy ($p_{i,t}^{PV}$) and load demands ($p_{i,t}^L$), to guarantee the minimal recourse control cost and the system security. The proposed model aims to enhance operational flexibility and coordination, ensuring a less-conservative strategy under actual operating conditions.

\subsubsection{Objective Function}
The recourse control cost $(\Lambda_2)$ is determined by the cost of recourse active power output of the DGs and the cost of additional electricity purchased from the main grid. 
\begin{equation} \label{obj2}
\min \Lambda_2 = \sum_{t \in {\cal T}} \left(\sum_{i \in {\cal G}} \hat{c}_i \hat{p}_{i,t}^G + \hat{d}_t (\hat{p}_t^{S} - p_t^{S})\right)
\end{equation}

\subsubsection{DGs Recourse Constraints}
The DGs constraints are composed of the recourse capability limits \eqref{eq:DGs_recourse}, the reactive power output limits \eqref{eq:DGs_reactive}, the active power output after recourse control \eqref{eq:DGs_regulate} as well as the ramp up/down rate limits of the active output after recourse control \eqref{eq:DGs_regulate_rampup}--\eqref{eq:DGs_regulate_rampdn}.
\begin{subequations}
\begin{align}
    \label{eq:DGs_recourse}
    &\underline{\hat{p}}_{i}^{G} \leq \hat{p}_{i,t}^G \leq \overline{\hat{p}}_{i}^{G}, \forall i \in {\cal G}, \forall t \in {\cal T} \\
    \label{eq:DGs_reactive}
    &\underline{\hat{q}}_{i}^{G} \leq \hat{q}_{i,t}^G \leq \overline{\hat{q}}_{i}^{G}, \forall i \in {\cal G}, \forall t \in {\cal T} \\
    \label{eq:DGs_regulate}
    &\widehat{p}_{i,t}^{G} =  p_{i,t}^{G} + \hat{p}_{i,t}^G, \forall i \in {\cal G}, \forall t \in {\cal T} \\
    \label{eq:DGs_regulate_rampup}
    &\widehat{p}_{i,t}^{G} - \widehat{p}_{i,t-1}^{G} \leq RU_{i}, \forall i \in {\cal G}, \forall t \in {\cal T} \\
    \label{eq:DGs_regulate_rampdn}
    &\widehat{p}_{i,t-1}^{G} - \widehat{p}_{i,t}^{G} \leq RD_{i}, \forall i \in {\cal G}, \forall t \in {\cal T}
\end{align}
\end{subequations}

\subsubsection{Power Flow and Voltage Constraints}
Constraints \eqref{eq:power_balance_2s}-\eqref{eq:reactivepower_balance_2s} model the active/reactive power balance equations. Constraints \eqref{eq:netload_2s}-\eqref{eq:reac_load_2s} consider the net active/reactive loads $(\hat{p}_{i,t}/\hat{q}_{i,t})$, and \eqref{eq:voltage_2s} is the voltage magnitude drop in the lines.
\begin{subequations}
\begin{align}
    \label{eq:power_balance_2s}
    &\sum_{hi \in \mathcal{L}} \hat{p}_{hi,t} - \sum_{ij \in \mathcal{L}}\hat{p}_{ij,t} = \hat{p}_{i,t}, \forall i \in \mathcal{N}, \forall t \in \mathcal{T} \\
    \label{eq:reactivepower_balance_2s}
    &\sum_{hi \in \mathcal{L}} \hat{q}_{hi,t} - \sum_{ij \in \mathcal{L}}\hat{q}_{ij,t} = \hat{q}_{i,t}, \forall i \in \mathcal{N}, \forall t \in \mathcal{T} \\
    \label{eq:netload_2s}
    &\hat{p}_{i,t} = \widehat{p}_{i,t}^{G} + p_{i,t}^{B+} - p_{i,t}^{B-} + p_{i,t}^{PV} - p_{i,t}^{L}, \forall i \in \mathcal{N}, \forall t \in \mathcal{T} \\
    \label{eq:reac_netload_2s}
    &\hat{q}_{i,t} = \hat{q}_{i,t}^{G} - q_{i,t}^{L}, \forall i \in \mathcal{N}, \forall t \in \mathcal{T} \\
    \label{eq:reac_load_2s}
    &q_{i,t}^{L} = p_{i,t}^{L} \tan\delta_{i,t}, \forall i \in \mathcal{N}, \forall t \in \mathcal{T} \\
    \label{eq:voltage_2s}
    &\hat{v}_{i,t} - \hat{v}_{j,t} = (r_{ij}\hat{p}_{ij,t} + x_{ij}\hat{q}_{ij,t})/(1-\varphi_{ij}), \forall ij \in \mathcal{L}, \forall t \in \mathcal{T}
\end{align}
\end{subequations}

The Lossy DistFlow formulation, with the help of parameter $\varphi_{ij}$ \cite{lossy_DistFlow}, is employed to approximate the loss terms in voltage magnitude drop calculations within the traditional DistFlow model, as shown in \eqref{eq:voltage_2s}. This approach linearizes the voltage calculations while preserving a degree of accuracy. 

\subsubsection{Network Security Constraints}

The maximum power flow limits and voltage magnitude limits are enforced by constraints \eqref{eq:power_limit_2s}-\eqref{voltage_limit_2s}. 
\begin{subequations}
\begin{align}
    \label{eq:power_limit_2s}
    &-\overline{p}^{l} \leq \hat{p}_{ij,t} \leq \overline{p}^{l}, \forall ij \in {\cal L}, \forall t \in {\cal T} \\
    \label{voltage_limit_2s}
    &\underline{v} \leq \hat{v}_{i,t} \leq \overline{v}, \forall i \in {\cal N}, \forall t \in {\cal T}
\end{align}
\end{subequations}

Given the uncertainties of RESs and load demands, integrating the economic dispatch plan with a recourse control stage enhances operational flexibility. Compared to a single-stage approach, this model allows for dynamic adjustments by making recourse decisions in response to actual variations in uncertainties, and co-optimizes active and reactive power dispatch across different timescales, thereby minimizing the need for conservative strategies and ensuring system security against uncertain scenarios \cite{two-stage_modeling_DRO}.

\section{Confidence Level-Based Information Gap Decision Formulation}
\label{CL-IGDT}

\subsection{Deterministic Model}
A compact matrix-vector form of the optimization model presented in Section~\ref{Mathematical Formulation} is provided below for subsequent reformulation purposes.
\begin{subequations}
\label{eq:deterministic_matrix}
\begin{align}
    \label{d1}
    \min_{\bm{x},\bm{y}} \quad & \Lambda(\bm{x},\bm{y}, \bm{\tilde{\xi}}) = \bm{c}^T \bm{x} + \bm{d}^T \bm{y} \\
    \label{d2}
    \text{s.t.} \quad & \bm{A} \bm{x} \leq \bm{b} \\
    \label{d3}
    & \bm{E} \bm{x} \leq \bm{F}\bm{\tilde{\xi}} \\
    \label{d4}  
    & \bm{G} \bm{x} + \bm{H} \bm{y} \leq \bm{0} \\
    \label{d5}
    & \bm{I} \bm{x} + \bm{J} \bm{y} \leq \bm{L} \bm{\tilde{\xi}} \\
    \label{d6} 
    & \bm{M} \bm{y} \leq \bm{g}
\end{align}
\end{subequations}

In the above deterministic formulation, the total operational cost is indicated by $\Lambda(\bm{x},\bm{y}, \bm{\tilde{\xi}})$, where $\bm{x}$ refers to the first-stage decision variable including $[p_{i,t}^{G}, p_{i,t}^{B+}, p_{i,t}^{B-}, z_{i,t}]$, $\bm{y}$ is the second-stage decision variable including $[\hat{p}_{i,t}^{G}, \hat{q}_{i,t}^{G}]$, and $\bm{\tilde{\xi}}=\bm{\tilde{p}^{L}}-\bm{\tilde{p}^{PV}}$ denotes the uncertain vector related to load demands and renewable energy, which adopt their forecasted values in the deterministic model. The variable $\bm{x}$ is constrained by \eqref{d2}, corresponding to constraints \eqref{eq:DGs}--\eqref{eq:ESSs}, \eqref{eq:power_balance}, and \eqref{eq:power_limit}. Constraint \eqref{d3} correlates $\bm{x}$ with the forecasted value of uncertain vector $\bm{\tilde{\xi}}$, corresponding to constraint \eqref{eq:netload}. Constraint \eqref{d4} builds the relationship between $\bm{x}$ and $\bm{y}$, corresponding to constraint \eqref{eq:DGs_regulate}. Constraint \eqref{d5} correlates $\bm{x}$ and $\bm{y}$ with the forecasted value of uncertain vector $\bm{\tilde{\xi}}$, corresponding to constraints \eqref{eq:netload_2s}--\eqref{eq:reac_load_2s}. Finally, the remaining constraints concerning $\bm{y}$ are represented by \eqref{d6}.

\subsection{Information-Gap Decision Model}

DNs face significant uncertainties associated with RESs and load demands, necessitating a framework for uncertainty management. IGDT offers a non-probabilistic, non-fuzzy alternative that effectively handles uncertainties without the need to predefine the uncertainty extent \cite{igdt_Ben_Haim}. The commonly used envelope-bound modelling by IGDT is shown below:

\begin{equation}
\bm{\mathit{\Psi}}(\bm{\tilde{\xi}}, \delta) = \left\{ \bm{\xi} \mid \left| (\bm{\xi} - \bm{\tilde{\xi}})\right| / \bm{\tilde{\xi}} \leq \delta, 0 \leq \delta \leq 1 \right\}
\end{equation}
where the variable $\delta$ represents the uncertainty extent.

\begin{figure}[H]
\centering
\includegraphics[width=0.5\linewidth]{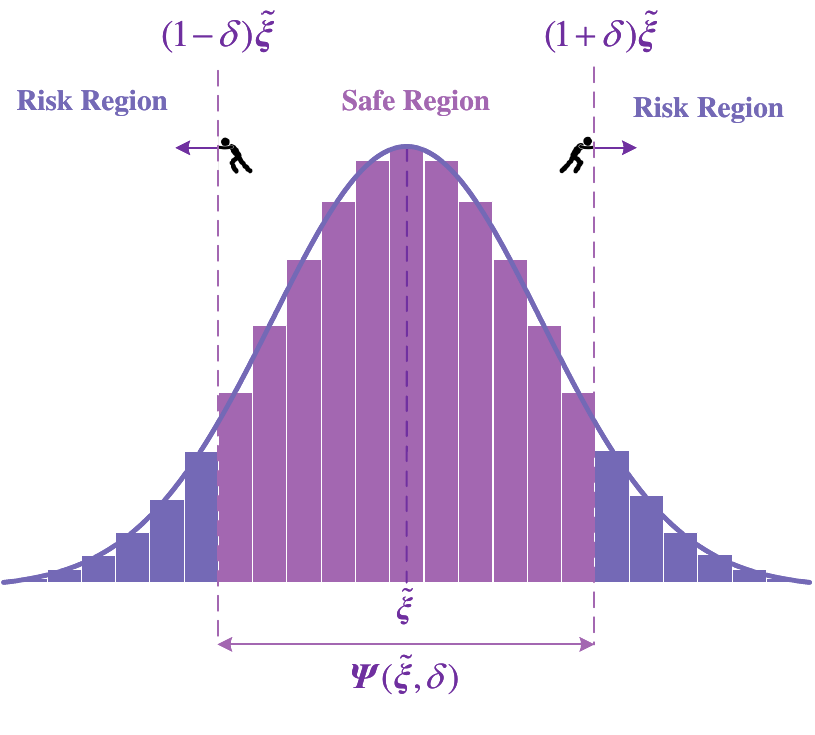}
\caption{The symmetric uncertainty set aligns perfectly with the shortest confidence interval of uncertain parameter characterized by symmetric probability density function.}
\label{fig:symmetric_distribution}
\end{figure}

The uncertainty modelling $\bm{\mathit{\Psi}}(\tilde{\bm{\xi}}, \delta)$ highlights the gap between the forecasted values and unknown information. For a given $\tilde{\bm{\xi}}$, an increasing $\delta$ enhances the risk-averse capability, leading to a more costly solution, as it requires more resources to ensure performance and reliability under uncertainty. A major achievement of IGDT is its risk-averse capability, which refers to the maximum uncertainty extent obtained by driving the actual system performance towards the desired target. As Fig.~\ref{fig:symmetric_distribution} shows, IGDT maximizes the uncertainty extent (safe region) to enable robust operational strategies. This is particularly advantageous in the energy dispatch field when clear performance targets, such as the operational budget, are set by the DSO. Under the desired target, the risk-averse capability is maximized. Consequently, this paper applies IGDT to handle the uncertain parameters in \eqref{eq:deterministic_matrix}, which can thus be reformulated as follows:
\begin{subequations}
\label{igdt}
\begin{align}
\max_{\bm{x},\bm{y}} \quad & \delta \label{igdt1} \\
\text{s.t.} \quad & \bm{A} \bm{x} \leq \bm{b} \label{igdt3} \\
& \bm{E} \bm{x} \leq \bm{F}\bm{\tilde{\xi}} \label{igdt4} \\
& \bm{G} \bm{x} + \bm{H} \bm{y}(\bm{\xi}) \leq \bm{0}, \forall \bm{\xi} \in \bm{\mathit{\Psi}}(\bm{\tilde{\xi}},\delta) \label{igdt5} \\
& \bm{I} \bm{x} + \bm{J} \bm{y}(\bm{\xi}) \leq \bm{L}\bm{\xi}, \forall \bm{\xi} \in \bm{\mathit{\Psi}}(\bm{\tilde{\xi}},\delta) \label{igdt6} \\
& \bm{M} \bm{y}(\bm{\xi}) \leq \bm{g}, \forall \bm{\xi} \in \bm{\mathit{\Psi}}(\bm{\tilde{\xi}},\delta) \label{igdt7} \\
& \overline{\Lambda} \leq (1+\sigma) \Lambda_{0} \label{igdt8} \\
& \overline{\Lambda} = \bm{c}^T \bm{x} + \max_{\bm{\xi} \in \bm{\mathit{\Psi}}(\bm{\tilde{\xi}}, \delta)} \bm{d}^T \bm{y}(\bm{\xi}) \label{igdt9}
\end{align}
\end{subequations}

IGDT formulation in \eqref{igdt} is a two-level optimization model. At the upper level, the objective is to maximize the uncertainty extent while enforcing the constraints \eqref{igdt3}--\eqref{igdt8}. The lower level is specified in \eqref{igdt8}--\eqref{igdt9}, where $\sigma$ represents the deviation factor that defines the acceptable degree of budget excess, and $\Lambda_{0}$ is the system performance under the expected scenario. Through this two-level framework, robust performance $\overline{\Lambda}$ is deliberately driven towards the desired budget $(1+\sigma) \Lambda_{0}$, ultimately enabling the optimal uncertainty extent.

\begin{figure}[H]
\centering
\includegraphics[width=0.6\linewidth]{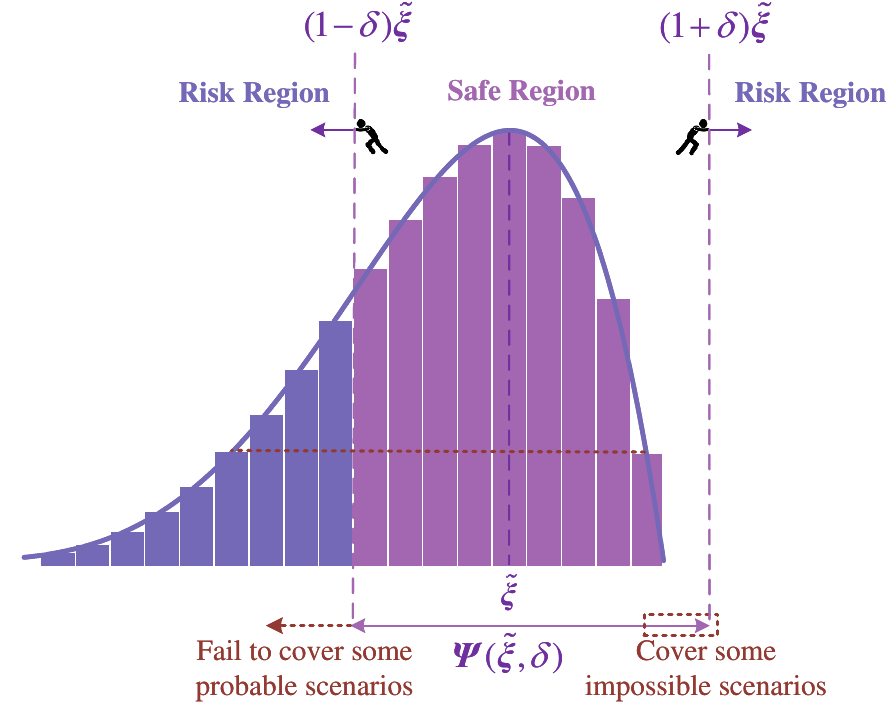}
\caption{For uncertain parameter with asymmetric characteristics, the symmetric uncertainty set no longer matches the shortest confidence interval.}
\label{fig:asymmetric_distribution}
\end{figure}

\begin{figure}[H]
\centering
\includegraphics[width=0.5\linewidth]{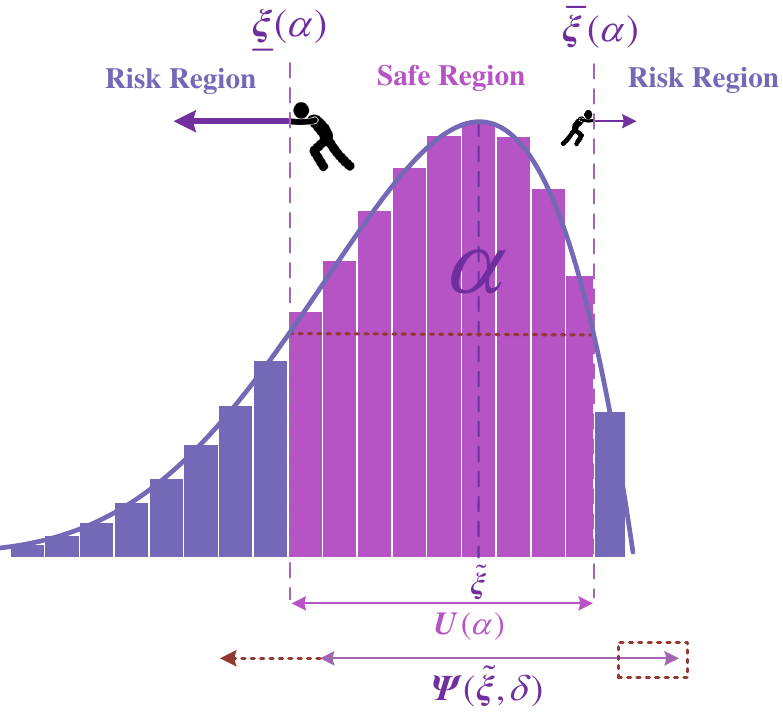}
\caption{The confidence level-driven uncertainty set captures the most probable uncertain scenarios, and thus matches the shortest confidence interval.}
\label{fig:asymmetric_set}
\end{figure}

The above uncertainty modelling requires little information on uncertainties and is well-suited for uncertain parameters exhibiting symmetric features. As shown in Fig.~\ref{fig:symmetric_distribution}, such uncertainty set can perfectly capture the scenarios most likely to occur. Uncertain parameters like renewable generation and load demands, however, typically do not follow a symmetric distribution \cite{asymmetric_load}. As demonstrated in Fig.~\ref{fig:asymmetric_distribution}, this misalignment can result in issues such as some safe regions being overlooked where actual scenarios would have occurred with higher probability. As mentioned in Section~\ref{Introduction}, these scenarios can compromise DNs safe operation if they occur. Additionally, the safe regions can be unnecessarily large due to the inclusion of impossible scenarios, leading to conservatism.

\subsection{Confidence Level Based Information Gap Decision Model} \label{subsec:CL-IGDT}
To mitigate the aforementioned drawbacks and better handle the asymmetric characteristics of uncertain parameters, we propose utilizing a confidence level-driven uncertainty set instead of the real-valued symmetric one. This approach transforms the uncertainty set into a sub-optimization problem:

\begin{align}
\label{eq:uncertainty set}
\begin{split}
    \bm{U}(\alpha) = \Bigg\{\bm{\xi} \mid \bm{\xi} \in \Big[ & \underline{\bm{\xi}}(\alpha), \overline{\bm{\xi}}(\alpha) \Big] 
    = \arg\min_{\overline{\bm{\xi}}, \underline{\bm{\xi}}} 
    \left\| \overline{\bm{\xi}} - \underline{\bm{\xi}} \right\|_{1} : \\
    & F_{\bm{\xi}}(\overline{\bm{\xi}}) - F_{\bm{\xi}}(\underline{\bm{\xi}}) \geq \alpha,\; 0 \leq \alpha \leq 1 
    \Bigg\}
\end{split}
\end{align}
\noindent where $\alpha$ represents the confidence level of the uncertainty set; $\overline{\bm{\xi}}(\alpha)$ and $\underline{\bm{\xi}}(\alpha)$ are the upper and lower bounds of the set, respectively, functions of $\alpha$; $F_{\bm{\xi}}(\cdot)$ denotes the cumulative distribution function (CDF) of $\bm{\xi}$.

As illustrated in Fig.~\ref{fig:asymmetric_set}, the newly constructed uncertainty set captures the scenarios that are most probable, corresponding to the shortest confidence interval. This achieves the highest robustness with the least conservatism and mitigates the misalignment issue. Accordingly, the objective of IGDT shifts to maximizing $\alpha$, allowing the uncertainty set to cover as many probable scenarios as possible within the desired budget. The improved IGDT formulation can be expressed as:
\begin{subequations}
\vspace{-0.5em}
\label{eq:cgdt}
\begin{align}
    \max_{\bm{x},\bm{y}} \quad & \alpha \label{cgdt1} \\
    \text{s.t.} \quad & \bm{A} \bm{x} \leq \bm{b} \label{cgdt3} \\
                      & \bm{E} \bm{x} \leq \bm{F}\bm{\tilde{\xi}} \label{cgdt4} \\
                      & \bm{G} \bm{x} + \bm{H} \bm{y}(\bm{\xi}) \leq \bm{0}, \forall \bm{\xi} \in \bm{U}(\alpha) \label{cgdt5} \\
                      & \bm{I} \bm{x} + \bm{J} \bm{y}(\bm{\xi}) \leq \bm{L}\bm{\xi}, \forall \bm{\xi} \in \bm{U}(\alpha) \label{cgdt6} \\
                      & \bm{M} \bm{y}(\bm{\xi}) \leq \bm{g}, \forall \bm{\xi} \in \bm{U}(\alpha) \label{cgdt7} \\
                      & \overline{\Lambda} \leq (1+\sigma) \Lambda_{0} \label{cgdt8} \\
                      & \begin{aligned}
                          \overline{\Lambda} &= \max_{\bm{\xi} \in \bm{U}(\alpha)} \Lambda \left( \bm{x},\bm{y}(\bm{\xi}), \bm{\xi} \right) \\
                                             &= \bm{c}^T \bm{x} + \max_{\bm{\xi} \in \bm{U}(\alpha)} \bm{d}^T \bm{y}(\bm{\xi})
                        \end{aligned} \label{cgdt9}
\end{align}
\end{subequations}

By analysing the formulation in Section \ref{Mathematical Formulation}, it is evident that the largest operational cost occurs when the electricity supply is at its peak, i.e., when renewable generation and load demands reach their bounds $\bm{\xi}^*(\alpha)=\overline{\bm{p}}^{L}(\alpha)-\underline{\bm{p}}^{PV}(\alpha)$ within the uncertainty sets. Consequently, the two-level optimization model can be simplified into a single level one if the relatively complete assumption holds. Additionally, in standard IGDT, a single variable $\delta$ becomes less practical when multiple types of uncertain parameters are involved. In contrast, the proposed model represents multiple uncertainties through their respective probability distributions. Although the sets are mapped from the same $\alpha$, each is linked to its own distribution, thereby achieving parameter decoupling compared to IGDT.

\subsection{Construction of the Ambiguity Set \label{subsec:ambiguity_set}}
Introducing a confidence level $\alpha$ in \eqref{eq:uncertainty set} and \eqref{eq:cgdt} necessitates the CDFs, which may inherit the limitations associated with SP method. To address this challenge, we draw inspiration from DRO and propose constructing ambiguity sets of CDFs. By fully utilizing available historical data and exploiting the implicit information it contains, the worst-case distribution can serve as a surrogate for the precise distribution.

Consider a dataset $\mathcal{D}$ of a random variable $\xi$ with $K$ different values $\xi_k, k=1,...,K$, arranged in an increasing order. The CDF of $\xi$ is defined as $F_{\xi}(x) = \Pr(\xi \leq x)$, so the cumulative probability of $\{\xi = \xi_k\}$ is $\theta_k = \Pr(\xi \leq \xi_k)$. According to the Law of Large Numbers, $\theta_k$ can be estimated as $n^{*}_k/n$ as the total sample counts $n \to \infty$, where $n^{*}_k$ is the cumulative counts corresponding to $\xi \leq \xi_k$. However, the limited size of $\mathcal{D}$ introduces inaccuracy into the probability estimation. To quantify the imprecision in $\theta_k$, the interval-valued probability is applied based on imprecise probability theory \cite{idm_P.Wally}, which can be estimated using credible index corresponding to IDM \cite{idm_Li_Peng}:
\begin{equation}
\label{eq:IDM_interval}
\begin{cases}
\underline{\theta}_k = 0, \overline{\theta}_k = W^{-1}\left(\frac{1+\gamma}{2}\right), \hfill n^{*}_k = 0 \\
\underline{\theta}_k = V^{-1}\left(\frac{1-\gamma}{2}\right), \overline{\theta}_k = W^{-1}\left(\frac{1+\gamma}{2}\right), \hfill 0 \leq n^{*}_k \leq n \\
\underline{\theta}_k = V^{-1}\left(\frac{1-\gamma}{2}\right),\overline{\theta}_k = 1, \hfill n^{*}_k = n
\end{cases}
\end{equation}
\noindent where $\gamma$ is the confidence index to estimate the probability interval, and $s$ is the equivalent sample size. $V(\cdot)$ and $W(\cdot)$ are the CDFs of Beta distribution, i.e., $\text{Beta}(n^{*}_k,s+n-n^{*}_k)$ and $\text{Beta}(s+n^{*}_k,n-n^{*}_k)$, respectively. 

Due to the limited sample points in $\mathcal{D}$, a stair-step interpolation method is applied to obtain the probability intervals at non-sample points, which can be expressed as:
\begin{equation}
\label{eq:interpolation}
\begin{cases}
\underline{\Pr}(\xi \leq \xi_{\kappa}) = \underline{\theta}_k, \hfill \kappa \in [k, k+1)\\
\overline{\Pr}(\xi \leq \xi_{\kappa}) = \overline{\theta}_{k+1}, \hfill \kappa \in (k, k+1]
\end{cases}
\end{equation}
\noindent where $\kappa$ indicates the non-sample points between two sample points $k$ and $k+1$.

Finally, the ambiguity set is constructed by connecting the upper and lower bounds of these probability intervals, which can be constructed as:
\begin{equation}
\mathcal{A} = \{ F_{\xi} \mid F_{\xi}(x) \in [\underline{\Pr}({\xi} \leq x), \overline{\Pr}({\xi} \leq x)] \}
\end{equation}

\subsection{Selection of the Worst-case Distribution} \label{subsec:worst-case_dis}
\begin{figure}[!t]
\centering
\includegraphics[width=3.5in]{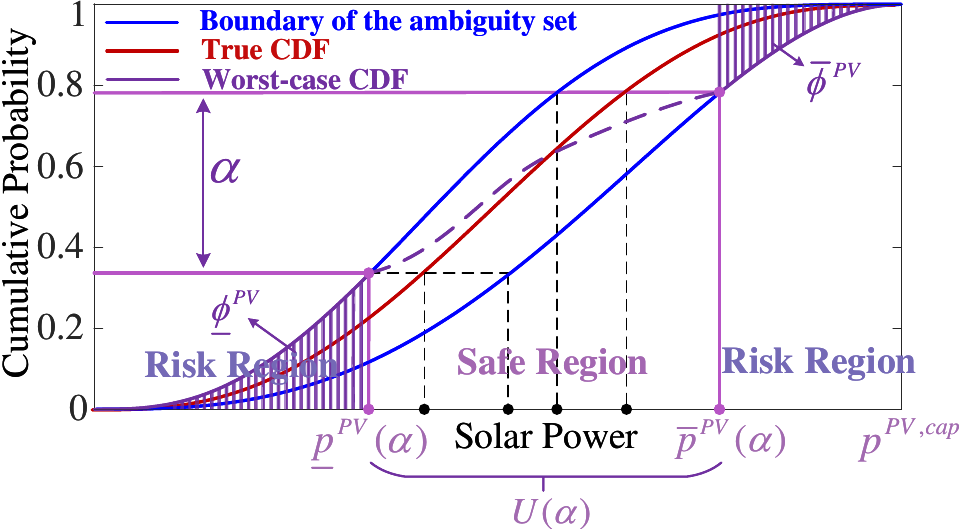}
\caption{Ambiguity set of solar power distribution. Given $\alpha$, the optimal uncertainty set over the worst-case distribution (illustrated with purple dots) is the widest compared to those derived from alternative distributions (represented by black dots). This suggests that alternative distributions, which result in narrower uncertainty sets and thus better economic performance, are not the worst-case distribution.}
\label{fig:ambiguity set}
\vspace{-1.0em}
\end{figure}

After counting the cumulative occurrences of different values of solar power $p^{PV}$, for example, \eqref{eq:IDM_interval} and \eqref{eq:interpolation} can be used to calculate the interval-valued cumulative probability and then construct the ambiguity set $\mathcal{A}_{PV}$, as depicted in Fig.~\ref{fig:ambiguity set}.

When actual solar power remains within the safe region, the total operational cost is constrained within the preset budget while ensuring network security. Conversely, disturbances that extend beyond this region into the risk areas may require additional actions to address unexpected conditions, or even potentially compromise the network security.

It is assumed that the upside risk is caused by excessive solar power, whereas the downside risk is due to insufficient solar power. The risks can be quantified based on the concept of expected risk described in \cite{idm_Li_Peng}.

The downside risk $(\underline{\phi}^{PV})$ is quantified as follows:
\vspace{-1em}
\begin{equation}
\begin{split}
\underline{\phi}^{PV}
&= \int_0^{\underline{p}^{PV}} (\underline{p}^{PV} - x) f_{p^{PV}}(x) \, dx \\
&= \int_0^{\underline{p}^{PV}} F_{p^{PV}}(x) \, dx
\end{split}
\end{equation}

Considering the ambiguity in CDFs, the maximal downside risk is:
\vspace{-1em}
\begin{equation}
\max_{F_{p^{PV}} \in \mathcal{A}_{PV}} \underline{\phi}^{PV} =\max_{F_{p^{PV}} \in \mathcal{A}_{PV}}\int_0^{\underline{p}^{PV}} F_{p^{PV}}(x) \, dx
\end{equation}

As illustrated in Fig.~\ref{fig:ambiguity set}, the maximal downside risk equals to the shaded area between $F_{p^{PV}}$ and the interval $[0,\underline{p}^{PV}]$. Obviously, $\overline{F}_{p^{PV}}$ leads to the highest downside risk, thus representing the worst-case distribution when the actual solar power exceeds its left boundary of the uncertainty set. Similarly, $\underline{F}_{p^{PV}}$ represents the worst-case distribution when the real solar power exceeds the right boundary. The worst-case distribution of solar power $F^{*}_{p^{PV}}(\cdot)$ can be expressed as:
\begin{equation}
    F^{*}_{p^{PV}}(x) = 
    \begin{cases} 
    \overline{F}_{p^{PV}}(x), & p^{PV} \leq \underline{p}^{PV} \\
    \underline{F}_{p^{PV}}(x), & p^{PV} \geq \overline{p}^{PV}
    \end{cases}
\end{equation}

With the construction of the ambiguity sets and the selection of the worst-case distribution pairs, the confidence level-driven uncertainty set in \eqref{eq:uncertainty set}, which was originally based on precise CDFs, is now replaced by the one constructed using ambiguity sets, as \eqref{eq:idm_uncertainty set} shows.
\begin{align}
\label{eq:idm_uncertainty set}
\begin{split}
    \bm{U}(&\alpha) = \Bigg\{ \bm{\xi} \mid 
    \bm{\xi} \in \Big[ \underline{\bm{\xi}}(\alpha), \overline{\bm{\xi}}(\alpha) \Big] 
    = \arg\min_{\overline{\bm{\xi}}, \underline{\bm{\xi}}} 
    \left\| \overline{\bm{\xi}} - \underline{\bm{\xi}} \right\|_{1} : \\
    &\underline{F}_{\bm{\xi}}(\overline{\bm{\xi}}) - \overline{F}_{\bm{\xi}}(\underline{\bm{\xi}}) \geq \alpha ,\;
    0 \leq \alpha \leq 1 \mid 
    \underline{F}_{\bm{\xi}}, \overline{F}_{\bm{\xi}} \in \mathcal{A} 
    \Bigg\}
\end{split}
\end{align}

\section{Solution Methodology} \label{Solution Methodology}
\begin{figure}[!t]
\centering
\includegraphics[width=3.5in]{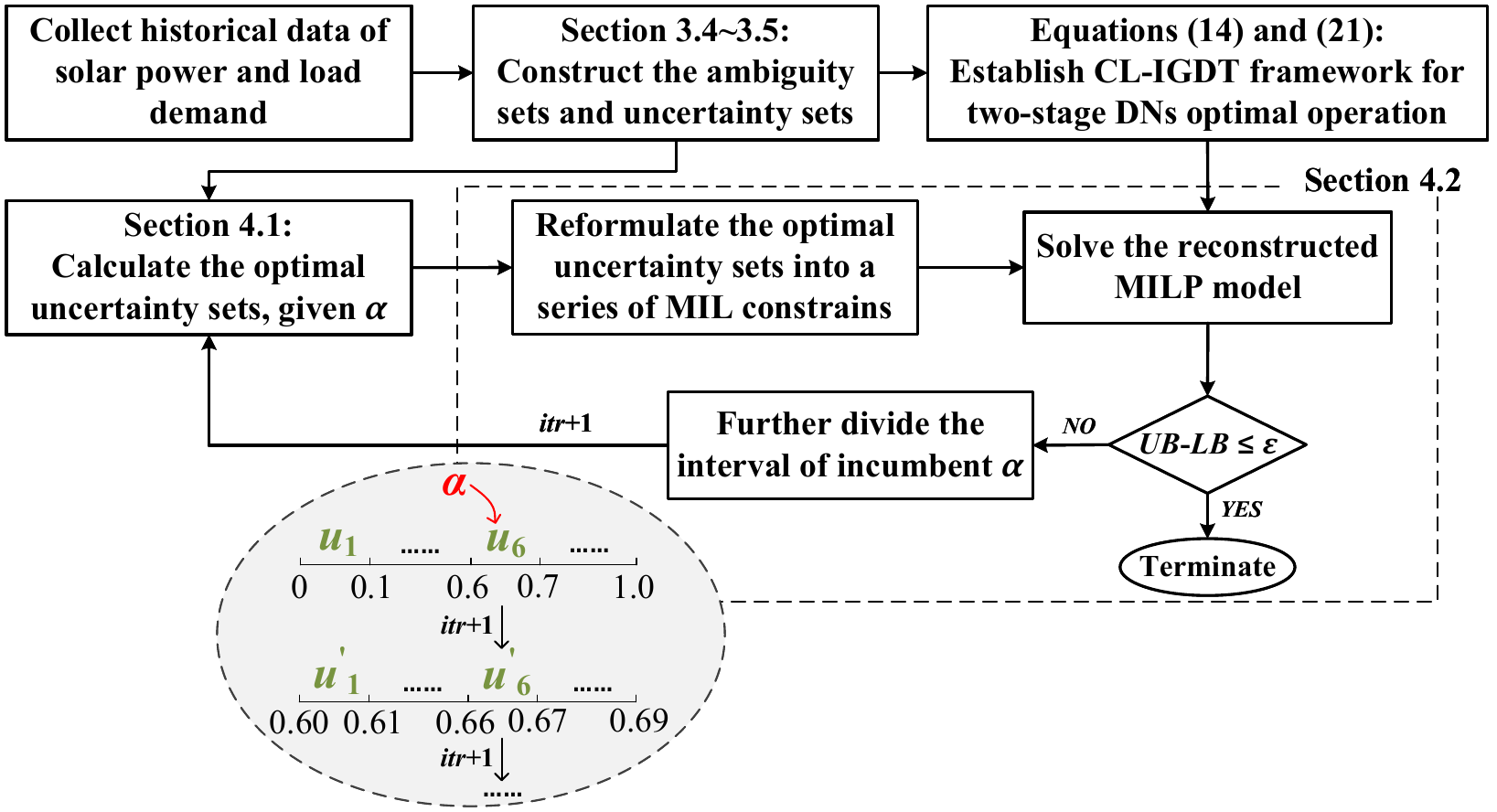}
\caption{Flowchart of the solution methodology.}
\label{fig:flowchart}
\end{figure}

This section presents the methodology to solve the proposed optimization framework shown in \eqref{eq:cgdt} and \eqref{eq:idm_uncertainty set}. An interval search procedure is carried out as a preprocessing step to identify the shortest confidence interval, thereby enabling the offline determination of the optimal uncertainty set. The optimization formulation is thus transformed into a mixed integer linear programming (MILP) model. An iterative method is introduced to determine the upper and lower bounds of the optimal $\alpha$ until the required accuracy is achieved. Overall, the proposed methodology is outlined in Fig.~\ref{fig:flowchart}.

\subsection{Calculation of Optimal Uncertainty Set} \label{subsec:Cal_Opti_Uset}
Explicitly expressing the uncertainty set in \eqref{eq:idm_uncertainty set} is challenging, as the suboptimization problem lacks closed-form distributions which are constructed by a series of discrete points, and thus deriving the optimality conditions is inapplicable. However, leveraging the structure of the proposed framework, an interval search approach is devised to calculate the uncertainty sets offline, which are then integrated into the original model. Again, we take solar power as an example. 

Consider the ambiguity set of $p^{PV}_{t}$ consists of $S$ ordered realizations. For each point $p^{PV}_{t,s}$ with $s=1,...,S$, its corresponding lower and upper bounds of the CDFs, $\underline{F}_{p^{PV}_{t}}(\cdot)$ and $\overline{F}_{p^{PV}_{t}}(\cdot)$, are available.

Assume $\alpha$ is given, all candidate intervals $[\underline{p}^{PV}_{t}, \overline{p}^{PV}_{t}]$ formed by the discrete points are examined by
\begin{equation}
\underline{F}_{p^{PV}_{t}}(\overline{p}^{PV}_{t}) - \overline{F}_{p^{PV}_{t}}(\underline{p}^{PV}_{t}) \ge \alpha \label{eq:cdf_inequality}
\end{equation}

Among all feasible intervals, the one with the minimum length is selected. This procedure yields the shortest confidence interval of solar power in \eqref{eq:idm_uncertainty set}, which can be computed offline prior to solving the optimization problem. The resolution of the resulting uncertainty set is determined by the number of available data points.

\subsection{Iteration Solving Procedure}
The optimal uncertainty set depends on the optimal $\alpha$, which remains unknown until the problem is solved. Fortunately, given the range of the objective function is from 0 to 1, the structured nature of the optimization problem can be exploited to allow for an iterative method, where a larger objective value will lead to a higher worst-case cost. The solving procedure is shown in Fig.~\ref{fig:flowchart}, and the details are as follows. 

\textit{Step 1:} The initial step is set as $itr=1$. Assume that $\alpha$ takes values from ten equal intervals separated by discrete points, $0,0.1,...,1.0$. Binary variables $u_m, m=1,...,10$ are utilized to determine which interval $\alpha$ falls into, as shown below.
\begin{equation}
\begin{cases}
    \alpha = \sum_{m=1}^{10} \alpha_{m} \\
    \underline{\alpha}_m^{(itr)} \cdot u_m \leq \alpha_{m} \leq \overline{\alpha}_m^{(itr)}  \cdot u_m, \forall m \\
    \sum_{m=1}^{10} u_{m} = 1
\end{cases}
\end{equation}
\noindent where
\begin{equation}
\begin{split}
    &\underline{\alpha}_m^{(itr)}=  10^{-itr}(m-1) \\
    &\overline{\alpha}_m^{(itr)} =  10^{-itr}m
\end{split}
\end{equation}

\textit{Step 2:} Given the values of each known $\underline{\alpha}_m$, the corresponding optimal uncertainty set of solar power and load demands can be obtained offline by using the approach in Section~\ref{subsec:Cal_Opti_Uset}, i.e., $U_{t}^{PV}(\underline{\alpha}_m) = [\underline{p}_{t}^{PV}(\underline{\alpha}_m),\overline{p}_{t}^{PV}(\underline{\alpha}_m)]$ and $U_{t}^{L}(\underline{\alpha}_m) = [\underline{p}_{t}^{L}(\underline{\alpha}_m),\overline{p}_{t}^{L}(\underline{\alpha}_m)], \forall t, \forall m = m \cup \{11\}$.

\textit{Step 3:} The optimal uncertainty sets corresponding to different $\underline{\alpha}_m$ are formulated as a series of constraints and then embed into the original optimization problem. The big M technique is utilized to determine which set of constraints associated with the optimal $\alpha$ are active, as shown below.
\begin{equation}
\begin{cases}
    \begin{aligned}
    \underline{p}_{t}^{PV}(\underline{\alpha}_m) &- M(1-u_m) \leq p_{t}^{PV} \\
    &\leq \overline{p}_{t}^{PV}(\underline{\alpha}_m) + M(1-u_m),
    \end{aligned} \\
    \begin{aligned}
    \underline{p}_{t}^{L}(\underline{\alpha}_m) &- M(1-u_m) \leq p_{t}^{L} \\
    &\leq \overline{p}_{t}^{L}(\underline{\alpha}_m) + M(1-u_m)
    \end{aligned}
\end{cases}
\end{equation}

\textit{Step 4:} The original problem now consists of a series of mixed integer linear constraints, transforming the entire optimization problem into a MILP, which can be solved directly by commercial solvers. Since $\alpha$ assumes taking value from only ten intervals, the incumbent solution may lack precision. The boundaries of the interval which the incumbent $\alpha$ falls into $[\underline{\alpha}^{(itr*)}, \overline{\alpha}^{(itr*)}]$ serve as the lower bound $LB$ and upper bound $UB$ of the optimal $\alpha^{*}$, respectively. To obtain more accurate result, we further divide the obtained interval into ten updated smaller intervals, and set the iteration $itr$ to $itr+1$.  
\begin{equation}
\begin{cases}
    \alpha = \sum_{m=1}^{10} \alpha_{m} \\
    (\underline{\alpha}^{(itr*)} + \underline{\alpha}_{m}^{(itr)}) \cdot u^{\prime}_m \leq \alpha_{m} \leq (\underline{\alpha}^{(itr*)} +  \overline{\alpha}_{m}^{(itr)}) \cdot u^{\prime}_m, \forall m \\
    \sum_{m=1}^{10} u^{\prime}_{m} = 1 
\end{cases}
\end{equation}

Given ten new discrete values, the corresponding optimal uncertainty set can be calculated offline again. Embed them into the original optimization problem and resolve it, so the new information can be updated as $UB=\overline{\alpha}^{(itr*)}$, $LB=\underline{\alpha}^{(itr*)}$, and $itr=itr+1$. 

\textit{Step 5:} Further segmentation of $\alpha$ into smaller intervals is possible if $\it{UB} - \it{LB} \leq \epsilon$ does not satisfy, where $\epsilon$ indicates the specified accuracy. \textit{Step 4} with smaller resolution will be repeated, or the optimal solution is assumed to be reached.

\section{Case Study and Discussion} \label{Case Study}
\begin{figure}[htbp]
\centering
\includegraphics[width=0.8\textwidth]{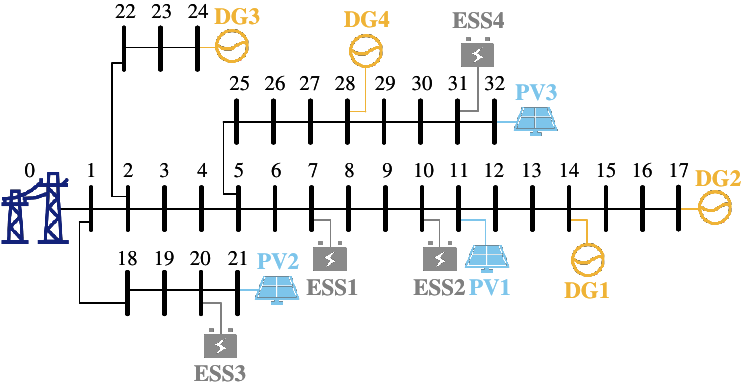}
\caption{Modified IEEE 33-bus system.}
\label{fig:IEEE33}
\end{figure}

The proposed framework was tested on the modified IEEE 33-bus system, as illustrated in Fig.~\ref{fig:IEEE33}. The system data is available in \cite{IEEE33}, and the data for DGs and ESSs is given in Table~\ref{tab:dg_data} and Table~\ref{tab:ESSs_data}, respectively. All simulations were conducted on a 64-bit PC with a 3.00 GHz CPU and 16 GB RAM in PyCharm. The simulation horizon was set to one day with a one-hour resolution. Solar power and load demand for each bus are considered uncertain sources. Approximately one year of historical data was used to construct the ambiguity sets. The confidence index $\gamma$ and equivalent sample size $s$ for IDM are set to 0.95 and 1, respectively \cite{idm_Li_Peng}.

\begin{table}[htbp]
\caption{DGs Information}
\vspace{0.5em} 
\label{tab:dg_data}
\centering
\footnotesize
\renewcommand{\arraystretch}{0.9}
\setlength{\tabcolsep}{4pt}
\begin{tabular}{lccccccc}
\toprule
\textbf{Label} & $\overline{p}^{G}$ (kW) & $\underline{p}^{G}$ (kW) & $RU$ (kW) & $RD$ (kW) & $\overline{q}^{G}$ (kVar) & $\underline{q}^{G}$ (kVar) & $c$ (\$) \\
\midrule
DG1 & 600 & 100 & 100 & 100 & 500 & -400 & 4.0 \\
DG2 & 400 & 100 & 90 & 90 & 300 & -200 & 3.5 \\
DG3 & 250 & 80 & 60 & 60 & 200 & -150 & 2.5 \\
DG4 & 50 & 10 & 15 & 15 & 40 & -25 & 2.0 \\
\bottomrule
\end{tabular}
\end{table}

\begin{table}[t]
\centering
\caption{ESSs Information}
\vspace{0.5em} 
\label{tab:ESSs_data}
\footnotesize 
\renewcommand{\arraystretch}{0.9}
\setlength{\tabcolsep}{4pt} 
\scalebox{0.9}{ 
\begin{tabular}{lccccc}
\toprule
Label & $\overline{p}^{B+}$ (kW) & $\overline{p}^{B-}$ (kW) & $\overline{e}$ (kW) & $\underline{e}$ (kW) & $\eta_{B}$ \\
\midrule
ESS1 & 120 & 120 & 600 & 120 & 0.9 \\
ESS2 & 40 & 40 & 200 & 40 & 0.9 \\
ESS3 & 60 & 60 & 300 & 60 & 0.9 \\
ESS4 & 80 & 80 & 400 & 80 & 0.9 \\
\bottomrule
\end{tabular}}
\end{table}

\subsection{Optimal Operation Results for Distribution Networks}
To illustrate the basic results of the proposed method, the deterministic (DT) and IGDT-based approaches are chosen as benchmarks. The deviation factor $\sigma$ is set as 30\%, and all other parameters remain consistent. The optimal cost of DT is 216,773\$, which determines the desired budget for IGDT and CL-IGDT. The optimal uncertainty extent obtained by IGDT is 0.183. This result indicates that, by implementing the optimal operational scheme, whenever the actual solar power and load demands fall within the safe regions where deviations are within 18.3\% of their predicted values, system safety is ensured and the total operational cost will remain within the preset budget, or otherwise no guarantees can be made. For CL-IGDT, the optimal probability is 0.665. The same explanation applies, but the safe regions are defined by the shortest confidence intervals corresponding to this probability.

\subsection{Ambiguity Sets within Proposed Framework} \label{subsec:AmbiguitySet}
\begin{figure}[htbp]
\centering
\includegraphics[width=0.7\textwidth]{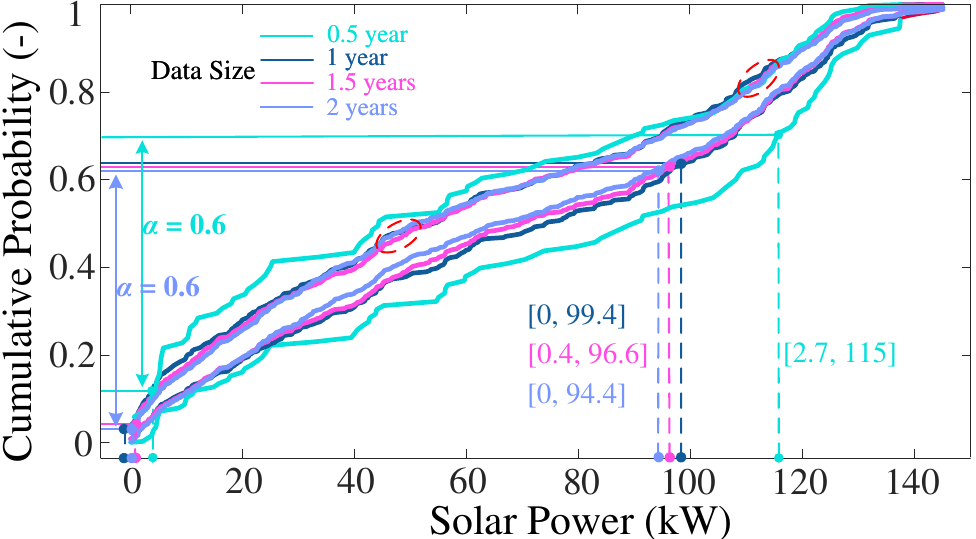}
\caption{Ambiguity sets of solar power distribution with different data sizes.}
\label{fig:ambiguity_set_pv}
\end{figure}

According to the previous research in \cite{idm_Li_Peng}, as the sample size consistently increases, the ambiguity set shrinks and the worst-case distribution is expected to be converged towards the true distribution. Consequently, the optimal uncertainty sets obtained should become progressively shorter, thereby reducing conservatism. Fig.~\ref{fig:ambiguity_set_pv} illustrates the ambiguity sets of solar power constructed using varying sizes of historical data and the optimal uncertainty set corresponding to each size, given $\alpha=0.6$. Interestingly, as the data size increases, the ambiguity set enlarges slightly in some areas (e.g., those highlighted in ellipses), and the optimal uncertainty sets do not completely envelop the one obtained from a larger data size, which appears to contradict the theory to some extent. Upon further analysis, two reasons for this contradiction emerge.

Firstly, the distribution's shape can initially be rough when the data size is small. This explains the significant change in the shape of the ambiguity set initially, but eventually stabilizing and approaching its real shape as more data is incorporated. Secondly, the initial sample size may be insufficient, and thus the limited discrete points inadequately represent the confidence level of 0.6, causing a deviation in the uncertainty set. It is suggested that with a sufficiently large dataset, the actual distribution and the optimal uncertainty set will be accurately captured, thereby reducing conservatism and aligning with theoretical expectations.

\subsection{Optimal Uncertainty Sets of Different Approaches} \label{subsec:OptimalUset}
\begin{figure*}[htbp]
\centering
\subfigure[]{\includegraphics[width=0.45\textwidth,keepaspectratio]{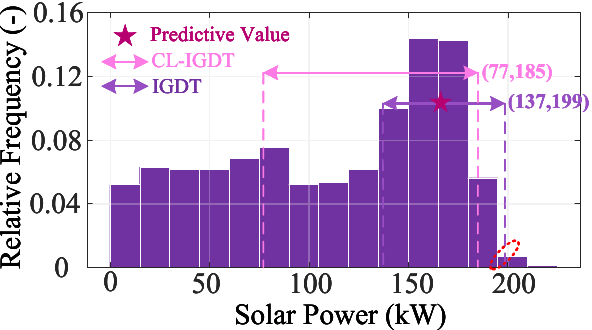}\label{fig:Opti_Uset_PV}}
\hfill
\subfigure[]{\includegraphics[width=0.45\textwidth,keepaspectratio]{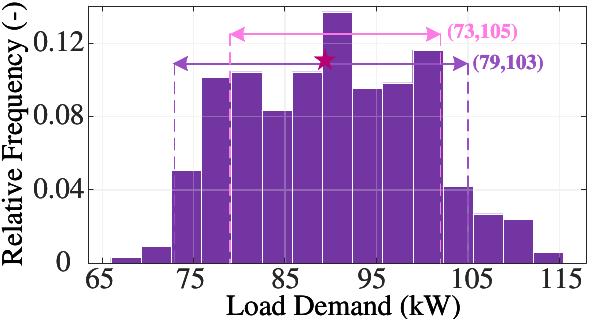}\label{fig:Opti_Uset_L}}

\vspace{1em} 

\subfigure[]{\includegraphics[width=0.45\textwidth,keepaspectratio]{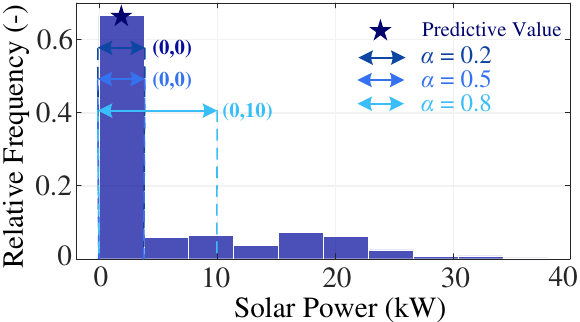}\label{fig:Uset_PV}}
\hfill
\subfigure[]{\includegraphics[width=0.45\textwidth,keepaspectratio]{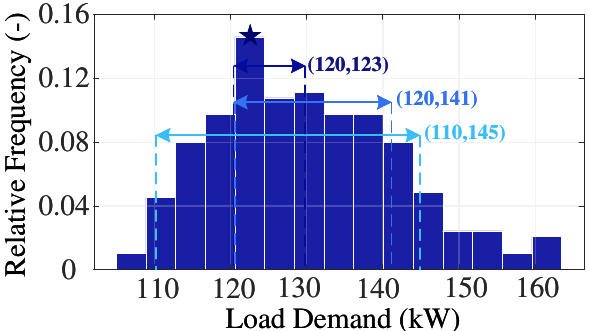}\label{fig:Uset_L}}
\caption{Optimal uncertainty sets. (a) Solar power for time period 11:00-12:00; (b) Load demand at node 20 for time period 1:00–2:00; (c) Solar power for time period 8:00-9:00; (d) Load demand at node 21 for time period 13:00–14:00.}
\label{fig:ex3_result}
\end{figure*}

As shown in Fig.~\ref{fig:Opti_Uset_PV}, the uncertainty set (safe region) obtained by CL-IGDT method is notably larger than that obtained by IGDT. Though larger uncertainty sets can lead to higher costs, it is probabilistically worthwhile to guarantee robustness at the expense of economic performance. It is because the safe region of the proposed method nearly (considering ambiguity) encompasses the scenarios with the highest probability of occurrence, whereas the uncertainty set obtained by IGDT is centered around its predictive value and includes some unnecessary scenarios (those highlighted in ellipse).

In contrast, Fig.~\ref{fig:Opti_Uset_L} illustrates the shorter uncertainty set obtained by CL-IGDT than IGDT. In this case, the most probable scenarios are relatively centered around its predictive value. To probabilistically ensure economic performance, scenarios with low probabilities of occurrence are excluded from the safe region to shorten the uncertainty set. It is also worth noting that the differences between the uncertainty sets obtained by the two approaches are reduced compared to Fig.~\ref{fig:Opti_Uset_PV}, as the asymmetry of the distribution is significantly mitigated. It is reasonable to infer that if the distribution exhibits symmetric characteristics, such as a Gaussian distribution, the difference in uncertainty set will diminish further, potentially becoming nearly identical. Therefore, the trade-off between economic performance (smaller uncertainty set) and robustness (larger uncertainty set) can be guaranteed in a probabilistic manner, depending on the feature of the data.

To further explore the superiority of the proposed uncertainty modelling, the optimal intervals under different $\alpha$ were examined. Fig.~\ref{fig:Uset_PV} shows that interval remains stable even though $\alpha$ increases from 0.2 to 0.5. While $\alpha$ reaches 0.8, this interval changes noticeably. The nonlinear expansion of the interval occurs since the solar power is predominantly centered around zero. For load demands without any distribution pattern shown in Fig.~\ref{fig:Uset_L}, the interval also expands nonlinearly rather than linearly as IGDT method. This indicates nonlinear robustness is considered in the proposed method.

\subsection{Post-Evaluation of Different Approaches}
The ‘here-and-now’ decision $\bm{x}$ is robustly optimized and serves as the final scheme, whereas the ‘wait-and-see’ decisions $\bm{y}$ can dynamically adjust in response to the realization of the uncertain scenarios. To assess the risk-averse capability of the CL-IGDT solution, post-evaluation is conducted on $\bm{x}$ obtained by different approaches, using 50 randomly selected scenarios which were not included in the ambiguity sets. Additionally, to evaluate the impact of historical data size on performance, two cases with limited data--one month and six months--are considered, denoted as CL-IGDT (limited-1) and CL-IGDT (limited-2), respectively. The performance metric is the expected project budget (EPB) \cite{EFB_CaoXY} defined in \eqref{eq:EPB}, and the reliability of the solutions, which refers to the percentage of feasible solutions.
\begin{equation} \label{eq:EPB}
EPB = \bm{c}^T\bm{x} + \frac{1}{NF} \sum_{n=1}^{NF} \bm{d}^T\bm{y}_n
\end{equation}
where $NF$ represents the number of common feasible solutions shared by different approaches.

\subsubsection{Considering Solar Power and Load Demand Uncertainties} 
\begin{figure}[htbp]
\centering
\subfigure[]{\includegraphics[width=0.6\linewidth,keepaspectratio]{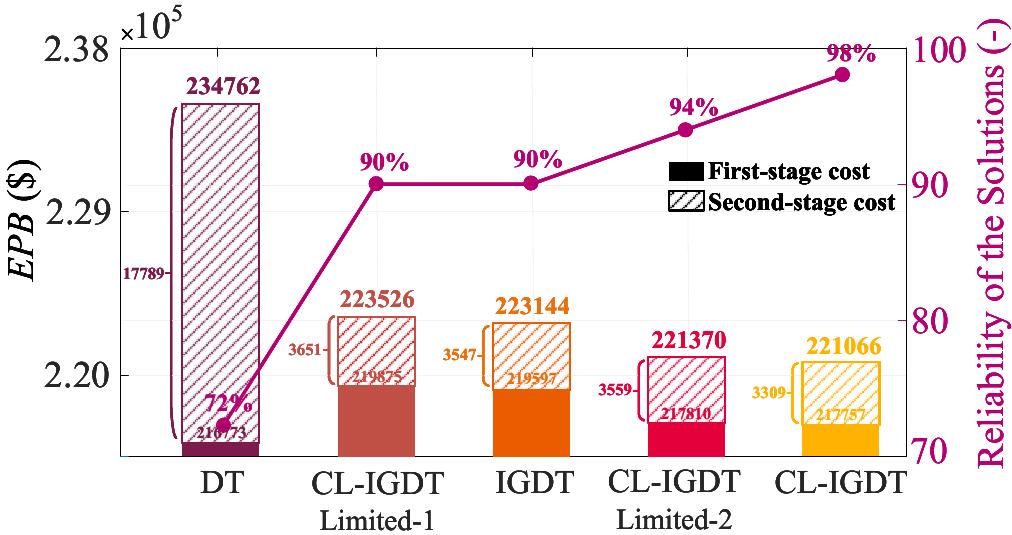}\label{fig:post_evaluation_1}}

\vspace{1em}

\subfigure[]{\includegraphics[width=0.6\linewidth,keepaspectratio]{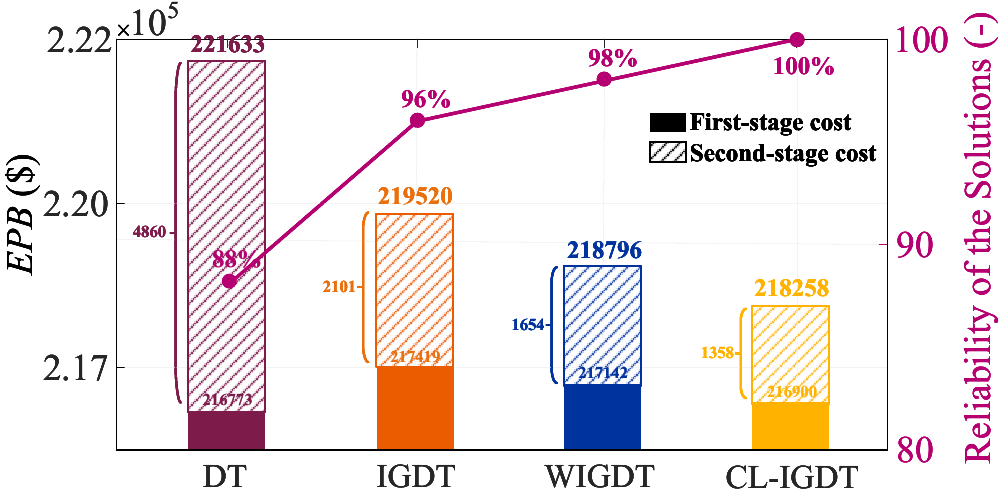}\label{fig:post_evaluation_2}}
\caption{Post-evaluation results by different approaches. (a) Considering solar power and load demand uncertainties; (b) Considering solar power uncertainty only.}
\label{fig:post_evaluation}
\end{figure}

As shown in Fig.~\ref{fig:post_evaluation_1}, the DT method incurs the highest \textit{EPB} and yields the lowest reliability. The optimal $\bm{x}^{*}$ contributes to the lowest first-stage cost because it is just optimized under expected scenarios. Without considering any potential uncertainty, the solution lacks risk-averse capability and requires substantial adjustments to compensate for the impact of the uncertain scenarios that were not taken into consideration during the decision process, thus resulting in an extremely high second-stage cost and the lowest reliability. In contrast, the robustness of the IGDT solution is achieved by investing 30\% more budget. By considering possible uncertain scenarios, higher first-stage cost is incurred but a significant reduction in the second-stage cost and a substantial increase in the reliability of the solutions are expected. 

The CL-IGDT approach further reduces \textit{EPB} and enhances reliability by accounting for the asymmetric feature of uncertainties and leveraging implicit patterns within the historical data to optimize $\bm{x}$. Compared to the IGDT, the proposed approach reduced the first-stage cost by 0.84\% and the second-stage average cost by 6.7\%. As discussed in Section~\ref{subsec:OptimalUset}, unnecessary robustness is avoided (which reduces the cost) while necessary robustness is maintained (which increases the cost). These two cases offset each other, so as to ensure the best trade-off between the risk-averse capability and economic performance (which can be demonstrated in Fig.~\ref{fig:post_evaluation_1} by the lower first-stage cost still contributing to lower second-stage cost and higher reliability). This explains why the improvement in economic performance, relative to the total operational costs, is relatively modest, while the reliability of the solutions is fairly enhanced by 8\% compared to IGDT. 

As data availability decreases, both economic performance and reliability of the proposed approach deteriorate. In particular, CL-IGDT (limited-1) incurs a \$382 higher \textit{EPB} than IGDT but with the same reliability percentage due to the lack of sufficient data resulting in rough ambiguity sets as discussed in Section~\ref{subsec:AmbiguitySet}. This proves IGDT’s superiority in handling severe uncertainties without any information. Given adequate data, the proposed approach achieves a better solution, and the superiority is expected to become more pronounced if the asymmetric characteristics of uncertainties become severer.

\subsubsection{Considering Solar Power Uncertainty} To ensure a fair comparison with an improved IGDT approach—WIGDT \cite{w_igdt}, which fails to handle multiple uncertainties, the load demands are fixed at their expected values and treated as deterministic parameters. As shown in Fig.~\ref{fig:post_evaluation_2}, the performance of WIGDT falls between IGDT and CL-IGDT. WIGDT accounts for the penetration levels of solar power by integrating them into the objective function as weights. Compared to IGDT, WIGDT offers greater flexibility, as it utilizes different variables to adjust uncertainty sets across different time periods. However, since WIGDT determines weights solely based on expected scenarios without leveraging historical solar power data, its performance remains inferior to the proposed method.

Additionally, a closer examination of Fig.~\ref{fig:post_evaluation_2} and Fig.~\ref{fig:post_evaluation_1} reveals that as multiple uncertainties are incorporated, the performance gap between IGDT and CL-IGDT widens, as reflected in the increase in economic performance difference from 0.57\% to 0.93\%, and in reliability difference from 4\% to 8\%. As discussed in Section~\ref{subsec:CL-IGDT}, IGDT relies on a single variable $\delta$ to control uncertainty sets for different types of uncertain parameters, limiting its flexibility. The $\alpha$ in proposed method models the risk-averse capability in a probabilistic manner, ensuring that the uncertainty sets for different uncertain parameters consistently cover the most probable scenarios.

\subsection{Sensitivity Analysis}
\begin{figure}[!t]
\centering
\includegraphics[width=2.5in]{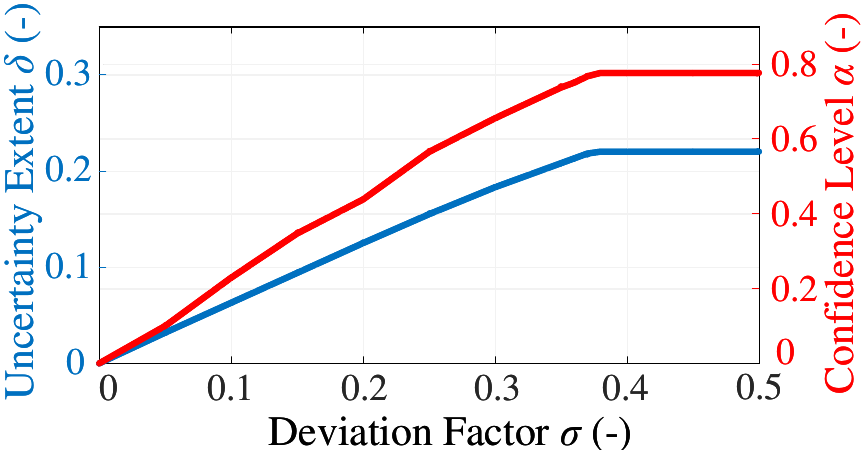}
\caption{Sensitivity analysis of deviation factor.}
\label{fig:sensitivity}
\end{figure}

The impact of the deviation factor $\sigma$ on the objective functions of IGDT and CL-IGDT methods is analyzed. Fig.~\ref{fig:sensitivity} illustrates their corresponding values of $\delta$ and $\alpha$, which indicate the risk-averse capability of the optimal solutions with regards to the varying $\sigma$. The risk-averse capability increases linearly as more budget is allocated, meaning that both approaches provide higher robustness at the expense of economical performance. However, as discussed in Section~\ref{subsec:OptimalUset}, the uncertainty sets of CL-IGDT increases nonlinearly even with a linear increase in $\alpha$, unlike the linear response of IGDT method. Notably, when $\sigma$ approaches around 0.4, the risk-averse capability plateaus for both approaches, signifying that the maximum level of robustness has been reached due to the limited availability of dispatchable resources within the DNs. This suggests that the risk-averse capability may further enhance if additional dispatchable resources are considered.

\subsection{Accuracy of the Optimal Solutions}
The iteration method in Section~\ref{Solution Methodology} is to determine the upper and lower bounds of the objective function in \eqref{cgdt1}. To test the validity of the proposed method, different accuracy requirements were tested. As iterations progress, the incumbent $\alpha$ becomes more accurate, as Table~\ref{tab:Accuracy} shows, thus leading to increasingly refined confidence level-driven uncertainty sets. Additionally, as the accuracy of $\alpha$ improves, the \textit{EPB} decreases and the reliability of the solutions is enhanced. This indicates that the incumbent solution progressively moves towards the optimal $\alpha^{*}$. Once the preset accuracy threshold is met, the incumbent solution is considered sufficiently optimal.
\begin{table}[t]
\centering
\caption{Accuracy of the optimal solutions}
\vspace{0.5em} 
\label{tab:Accuracy}
\footnotesize
\renewcommand{\arraystretch}{0.9} 
\setlength{\tabcolsep}{4pt}
\scalebox{0.9}{ 
\begin{tabular}{lccccc}
\toprule
Iterations & Incumbent $\alpha$ & Range of $\alpha^{*}$ & \textit{EPB} (\$) & Reliability & Time (s) \\
\midrule
1 & 0.6 & [0.60, 0.69] & 223453 & 94\% & 2.4 \\
2 & 0.66 & [0.660, 0.669] & 223434 & 98\% & 5.3 \\
3 & 0.665 & [0.6650, 0.6659] & 223421 & 98\% & 7.5 \\
4 & 0.6651 & [0.66510, 0.66519] & 223404 & 98\% & 10.1 \\
\bottomrule
\end{tabular}}
\end{table}

\section{Conclusion} \label{Conclusion}
This paper proposes a CL-IGDT framework for the two-stage optimal operation of distribution networks, addressing renewable energy and demand asymmetric uncertainties. Results show that with sufficient historical data, the proposed method achieves the best trade-off between economic performance and robustness, reducing first-stage cost by 0.84\%, second-stage average cost by 6.7\%, and improving solution reliability by 8\%. Otherwise, with insufficient data, confidence level-driven uncertainty sets become inaccurate, leading to performance deterioration. In such case, IGDT is preferable for handling severe uncertainties without prior information. As additional budget is allocated, the risk-averse capability increases linearly (while uncertainty sets expand nonlinearly) but finally plateaus due to limited dispatchable resources. The proposed methodology is validated by its ability to achieve lower \textit{EPB} and higher reliability over iterations.

\bibliographystyle{elsarticle-num}
\bibliography{distributionally_IGDT}

\end{document}